# Multiple scattering effect on angular distribution and polarization of radiation by relativistic electrons in a thin crystal

A.S. Fomin[*][a], S.P. Fomin[b], N.F. Shul'ga[b]

[a]Kharkiv National University, Phys.-Tech. Dep., 31 Kurchatov Ave., 61108 Kharkov, Ukraine
[b]Kharkiv Institute of Physics and Technology, 1 Akademicheskaya Str., 61108 Kharkov,Ukraine

(Dated: November 2, 2004)

## ABSTRACT

The multiple scattering of ultra relativistic electrons in an amorphous matter leads to the suppression of the soft part of radiation spectrum (the Landau-Pomeranchuk-Migdal effect), and also can change essentially the angular distribution of the emitted photons. A similar effect must take place in a crystal for the coherent radiation of relativistic electron. The results of the theoretical investigation of angular distributions and polarization of radiation by a relativistic electron passing through a thin (in comparison with a coherence length) crystal at a small angle to the crystal axis are presented. The electron trajectories in crystal were simulated using the binary collision model which takes into account both coherent and incoherent effects at scattering. The angular distribution of radiation and polarization were calculated as a sum of radiation from each electron. It is shown that there are nontrivial angular distributions of the emitted photons and their polarization that are connected to the superposition of the coherent scattering of electrons by atomic rows ("doughnut scattering" effect) and the suppression of radiation (similar to the Landau-Pomeranchuk-Migdal effect in an amorphous matter). It is also shown that circular polarization of radiation in the considered case is identically zero.

## 1. INTRODUCTION

The radiation process of a relativistic electron develops in a large spatial region along the direction of particle motion, which is called the coherence length [1]: $l_c \approx 2\gamma^2/\omega$, where $\gamma$ is the Lorenz-factor of the particle, $\omega$ is the energy of emitted photon (we use here the units system $\hbar = c = 1$). This length grows fast with electron energy increasing and with decreasing of photon energy $\omega$.

Landau and Pomeranchuk showed [2] that if in limits of coherence length of radiation an electron interacts with a large number of medium atoms, the multiple scattering of particle on these atoms could conduce to suppression of bremsstrahlung in comparison with the prediction of the Bethe-Heitler theory [3]. The effect appears when the root-mean-square angle of the electron multiple scattering $\vartheta_e$ during one coherence length of bremsstrahlung $l_c$ exceeds the characteristic angle of radiation of a relativistic particle $\theta \sim \gamma^{-1}$. The quantitative theory of the effect in an amorphous medium was offered by Migdal using the kinetic equation method [4]. That is why this effect is often called the Landau-Pomeranchuk-Migdal effect (or the LPM effect in abbreviated form).

The circumstantial experimental investigation of the LPM effect was carried out recently at SLAC [5] at electron energies 8 and 25 GeV. During this experiment there was measured the spectral density of electron radiation in a range of photon energies from parts of MeV up to several hundreds MeV for targets of a series of the elements from carbon up to uranium. The thickness of targets varied from parts of percents up to several percents of radiation length. The analysis of the data obtained showed good agreement between the predictions of the Migdal theory of the LPM effect and the experiment for relatively thick targets, but "unexpected" behavior of the radiation spectrum at low frequencies for thin targets [6].

---

[*] e-mail: fomax.ua@gmail.com

The SLAC experiment stimulated a new wave of theoretical investigations of the LPM effect. In [7] it was shown that observed in [6] deviations from predictions of the Migdal theory take place in the special case, when the target thickness was small in comparison with a coherence length of radiation $T < l_c$ whereas the Migdal theory describes the LPM effect in a boundless amorphous medium. The suppression of radiation due to the multiple scattering in a thin layer of matter was considered earlier in [8,9], however, the asymptotic formulas obtained there can't be used immediately for describing the SLAC experimental data. The quantitative theory of this effect was developed recently in [10-15] using different approaches of classical and quantum electrodynamics. The results obtained in these works are in a good agreement with the SLAC experimental data.

It was shown in [9,17,18] that the suppression effect of radiation similar to the LPM effect in an amorphous medium must also take place when relativistic electrons pass through a crystal under a small angle $\psi$ to the crystal axis. However, in this case we have the suppression of the coherent radiation of relativistic electrons in a crystal [1]. Due to the coherent effect in scattering of fast electrons by the atomic chains in a crystal ("doughnut scattering" effect) the root-mean-square angle of the multiple scattering $\bar{\vartheta}_{cr}$ can exceed significantly the corresponding value for an amorphous target of the same thickness [19]. That is why the condition of the LPM effect in a crystal can be fulfilled at a smaller energy of the projectile particles than in an amorphous matter. The efficiency of multiple scattering in a crystal can be controlled by changing the orientation angle $\psi$ instead of changing thickness as in the amorphous target case. This is the additional useful possibility for an experimental investigation of the LPM effect in a crystal [20].

The experimental study of the LPM effect in a crystal was carried out at CERN [21] in the middle of the 80th for electron and positron energies up to 20 GeV, and the results obtained there for relatively thick crystals were in a good agreement with the theoretical prediction [22]. Recently these experiments were resumed in CERN [23] for electron energies up to 243 GeV.

All these experimental and theoretical investigations were devoted to the analysis of the multiple scattering effect on radiation spectrum. However, the multiple scattering can influence essentially not only radiation spectrum but also the angular distribution of radiation as it was shown recently in [16] for a thin amorphous target case.

The present work is devoted to the theoretical investigation of multiple scattering effect on angular distributions and polarization of radiation by relativistic electrons passing through a thin crystal at a small angle to the crystal axis. The main attention is paid to the manifestation of the LPM effect in a thin crystal. Special character of the multiple scattering of relativistic electrons by atomic chains of a crystal ("doughnut scattering" effect) leads to the certain linear [24] and circular [25] polarizations of an accompanied coherent radiation. Taking into account the features of angular distributions of emitted photons and their polarization one can obtain a rather high degree of linear polarization even under the LPM effect condition using the optimal photon collimator. It is also shown that circular polarization of radiation for each electron in a thin crystal ($T \ll l_c$) is identically zero contrary to the thick crystal case [25].

## 2. GENERAL FORMULAS

The general condition of the LPM effect which coincides with the condition of non-dipole regime of radiation

$$\vartheta_e > \gamma^{-1} \qquad (1)$$

can be fulfilled first of all for relatively low energy region of emitted photons $\omega \ll \varepsilon$ so as to provide a long enough coherence length $l_c$. In this case we can neglect the quantum recoil effect at radiation and use formulas of classical electrodynamics.

The spectral-angular density of radiation by an electron of trajectory $\vec{r}(t)$ is determined in classical electrodynamics by the expression [26]

$$\frac{d^2 E}{d\omega do} = \frac{e^2}{4\pi^2} \left[\vec{k} \times \vec{I}\right]^2, \qquad \vec{I} = \int_{-\infty}^{\infty} \vec{v}(t)\, e^{i(\omega t - \vec{k}\vec{r})} dt \qquad (2)$$

where $\vec{k}$ and $\omega$ are the wave vector and the frequency of the radiated wave.

If the coherence length of radiation process is big in comparison with the thickness of the target $l_c \gg T$ then $\vec{I}$ can be represented as [9,26]

$$\vec{I} \approx \frac{i}{\omega}\left(\frac{\vec{v}'}{1-\vec{n}\vec{v}'} - \frac{\vec{v}}{1-\vec{n}\vec{v}}\right), \tag{3}$$

where $\vec{v}$ and $\vec{v}'$ are the electron velocities before and after scattering, $\vec{n} = \vec{k}/\omega$.

The general expression for the polarization matrix of radiation is

$$J_{ik} = \frac{e^2\omega^2}{4\pi^2}\left(\vec{e}_i\vec{I}\right)\left(\vec{e}_k\vec{I}^*\right), \tag{4}$$

where $\vec{e}_{i,k}$ are the polarization vectors which are the unit vectors orthogonal to the wave vector $\vec{k}$ and to each other: $\vec{e}_{i,k}\vec{n} = 0$, $\vec{e}_i\vec{e}_k = \delta_{ik}$.

The degree of the linear polarization of radiation is determined by expression

$$P = \frac{J_{11} - J_{22}}{J_{11} + J_{22}}. \tag{5}$$

The degree of the circular polarization of radiation is

$$P_{circular} = \frac{J_{12} - J_{21}}{J_{11} + J_{22}}. \tag{6}$$

In [25] there was proposed a method for producing high energy gamma-quanta with circular polarization by coherent radiation of ultrarelativistic electrons passing through a crystal at a small angle to the crystal axis. This method is based on specific features of non-dipole regime of radiation that allows obtaining a high enough degree (about 50 %) of circular polarization of emitted gamma-quanta using the special type of photon collimation. The natural limitation of efficiency of this method for hard part of the radiation spectrum is connected with rapid decreasing of the number of emitted gamma-quanta with photon energy increasing $dN/d\omega \sim 1/\omega$. However, there is one more effect that suppresses the emission of circular polarization photons in a soft part of the spectrum despite the increasing of the total number of emitted photons.

Really, if the energy of the emitted photon is small enough, so that the coherence length of radiation process is larger than crystal thickness ($l_c \gg T$) then the polarization matrix (4) in this approximation becomes symmetrical

$$J_{ik} = J_{ki}$$

and the degree of circular polarization of each emitted gamma-quanta (6) is equal to zero.

This statement can be formulated as the following general theorem: if the coherence length of radiation process is much larger than the effective spatial region in which a relativistic electron interacts with an external field, then the circular polarization of the emitted photon is identically zero.

The spectral-angular density of radiation can be written in terms of polarization matrix as

$$\frac{d^2E}{d\omega do} = J_{11} + J_{22}. \tag{7}$$

If we are interested in the angular distribution of radiation from an electron beam passing through a thin target, then the formula (7) is necessary for averaging over the scattering angles of the particles in matter. If the distribution function of the scattered particles $f(\vec{\vartheta}_e)$ is known, then the average value of spectral-angular density of radiation will be determined by the expression

$$\left\langle\frac{d^2E}{d\omega do}\right\rangle = \int d\vec{\vartheta}_e\, f(\vec{\vartheta}_e)\frac{d^2E}{d\omega do}. \tag{8}$$

Note that formula (8) is applicable to any targets. It is only required that the target thickness is small in comparison with the coherence length of radiation. The different characteristics of the scatterer will be exhibited only by the definite kinds of distribution function $f(\vec{\vartheta}_e)$.

For the amorphous target case the distribution function $f(\vec{\vartheta}_e)$ is determined by the Bethe-Molière function [26]. The multiple scattering effect on the spectral-angular distribution of the radiation by relativistic electrons in a thin amorphous target was studied in [16]. It was shown that in contrast with the Bethe-Heitler theory prediction [3] there is the minimum in the angular distribution of the emitted gamma-quanta in the initial direction of the electron beam when the condition of the LPM effect $\vartheta_e > \gamma^{-1}$ is fulfilled (see Fig. 2 in [16]).

If a beam of relativistic electrons passes through a crystal at a small angle ψ to one of crystallographic axes (axis z) there takes place a coherent effect in electron scattering, exhibited as a characteristic annular angular distributions of the particles outgoing from the crystal ("doughnut scattering" effect [19, 26]). This coherent effect takes place only for the scattering over azimuthal angle φ (see Fig.1) as a result of correlations in sequent scattering of a fast electron by atoms located along this crystal axis. In this case the magnitude of the root-mean-square angle of multiple scattering of electron can exceed substantially (by several times) the corresponding parameter for the electron scattering in an amorphous target of the same thickness [19], and the smaller is the target thickness, the greater is this difference.

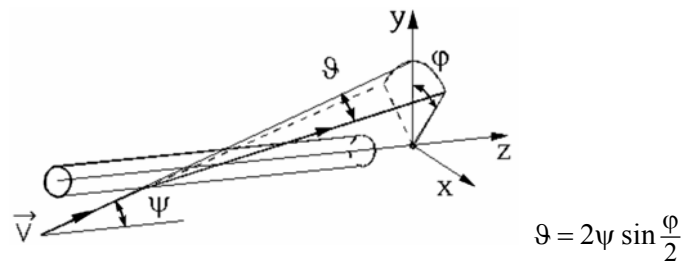

$$\vartheta = 2\psi \sin\frac{\varphi}{2}$$

Fig. 1. Geometry of the coherent azimuthal scattering in a crystal.

Generally the dynamics of a relativistic particle beam in an aligned crystal is rather complicated, since various fractions of a beam are involved in various regimes of motion: finite and infinite, regular and chaotic, with transitions between them. The analytical description of the particle dynamics can be conducted only in some limiting cases. Thus, for example, the theory of multiple scattering of relativistic charged particles on atomic strings of a crystal, based on the continuous string approximation, describes the coherent azimuthal scattering of above-barrier electrons [19]. However, this theory does not describe transitions of particles between two different fractions of the electron beam in the crystal, since the continuous string approximation does not take into account incoherent scattering. It is possible to take incoherent scattering into account by analytical methods only in the case of rather large incident angles $\psi \gg \psi_L$, where $\psi_L = \sqrt{4Ze^2/\varepsilon d}$ is the Lindhard angle [26], $Z|e|$ is the charge of atomic nuclear, d is the distance between atoms along the crystal axis and ε is the energy of the incident electron. At the same time, as it was already mentioned, the orientation effects in scattering and radiation of a relativistic electron beam passing through a crystal are mostly manifest in the range of angles $\psi \approx \psi_L$. Therefore, for the quantitative description of these effects, a computer simulation of the passing of an electron beam through an aligned crystal appears to be the most adequate.

### 3. RESULTS OF COMPUTER SIMULATION

With the purpose of a quantitative analysis of the multiple scattering effect on coherent radiation of a relativistic electron in a thin crystal, we performed a computer simulation on the basis of the Monte-Carlo method. We used here the binary collisions model of the electron interactions with the atoms of a crystalline lattice (see [20] and references therein). Such an approach allows taking into account both the coherent scattering of fast electrons on the atomic strings of a crystal and the incoherent scattering of the electrons connected with the thermal fluctuations of the atom positions in the lattice and with the electronic subsystem of the crystal. The angular distribution of radiation and polarization was calculated as a sum of radiation and polarization of each electron. The results of the computer simulation of electron scattering, radiation and polarization are presented in Figs. 2 and 3.

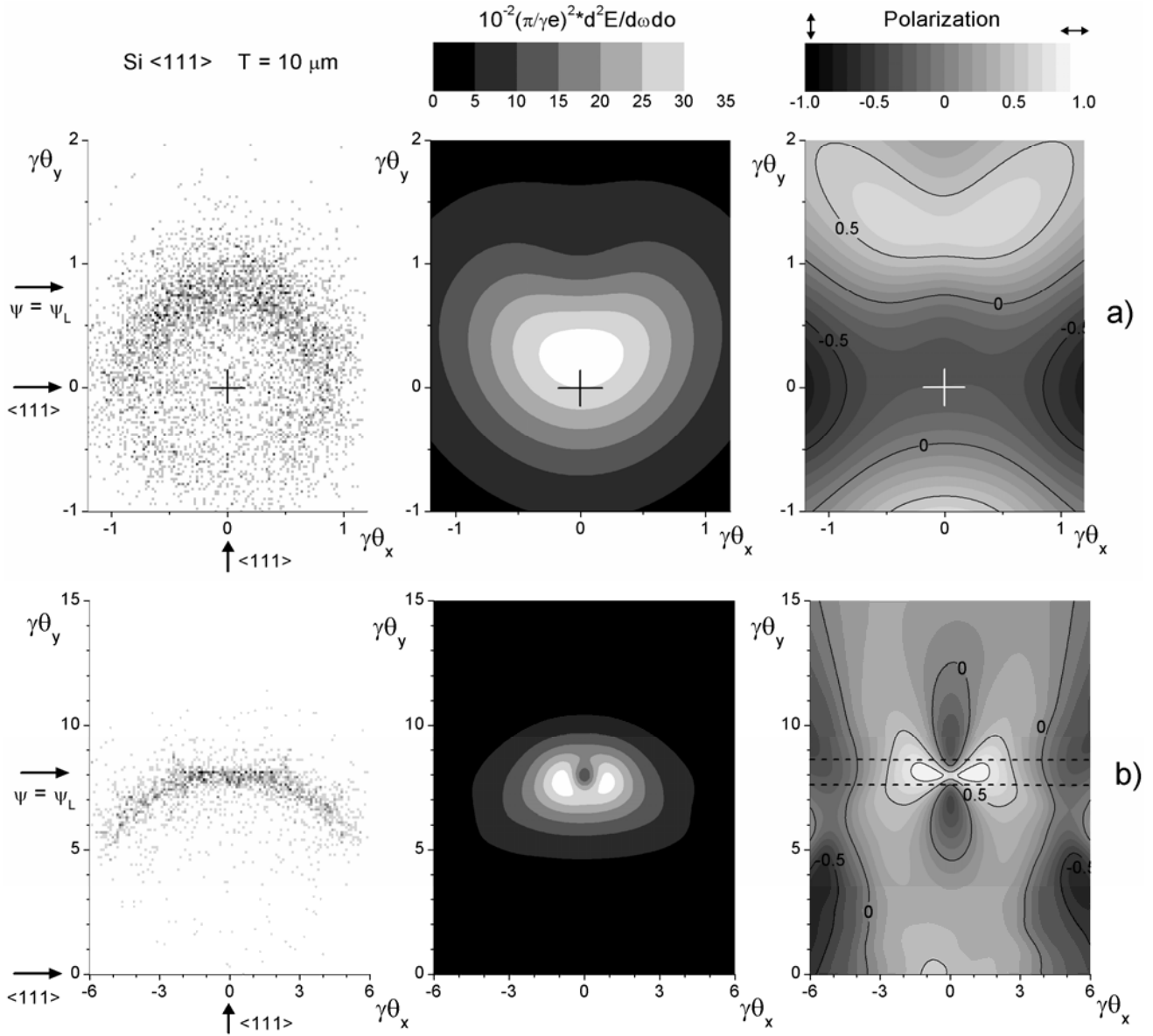

Fig. 2. The angular distributions of scattered electrons (left) as a result of computer simulation of 1 GeV (a) and 100 GeV (b) electron beam passed through the 10 μm silicon crystal when the electron beam is directed to the crystal at the angle $\psi = \psi_L$ to the axis <111>. Arrows show the direction of the crystal axis <111> and the initial electron beam direction. The angular distributions of intensity (middle) and linear polarization degree (right) of radiation by scattered electrons for 1 GeV (a) and 100 GeV (b) of the initial electron beam energy. Dashed lines shows the optimal position of the slit photon collimator with the angular width $\gamma^{-1}$ to obtain the highest degree of linear polarization.

The left part of Fig.2a represents the results of computer simulation of the angular distribution of 1 GeV electrons scattered by 10 μm silicon crystal when the electron beam is incident at the angle $\psi = \psi_L$ to the crystal axis <111>. The initial divergence of the electron beam was taken as 10 μrad that is equal to 0.02 in the units of $\gamma^{-1}$. Fig.2a (left) demonstrate the typical for the "doughnut scattering" effect annular angular distribution of scattered electron. The initial direction of the electron beam is shown in this figure by the arrow "$\psi = \psi_L$". The position of the crystal axis <111> is shown in Fig.2a by black (or white) crosses.

6The coherent azimuthal electron scattering for these initial conditions (electron beam energy $\varepsilon = 1$ GeV, incident angle $\psi = \psi_L$ and crystal thickness $T = 10$ μm) is strong enough to close the ring of the electron angular distribution around the crystal axis <111>. Incoherent scattering of the projectile relativistic electron on thermo vibrations of lattice atoms and on the electron subsystem of the crystal is isotropic in transverse plane (x,y) and it leads to the blooming of the coherent scattering ring. So, we can say that due to the "doughnut scattering" effect the relativistic electron beam is turned as a whole from the initial incident direction (that was $\psi = \psi_L$) to the crystal axis <111> with essential expansion of its angular size.

The middle part of Fig.2a represents the results of calculation of the angular distribution of radiation which is accompanied by the electron beam pass through the crystal. We use here the units $\gamma^{-1}$ that is a natural scale for angular distributions of relativistic particle radiation. The calculation was carried out using formula (8) in which the integration over the scattering angle $\vartheta_e$ was replaced by direct summation of contribution from each electron, scattered by the crystal at the angle $\vartheta_e$. The scattering angle $\vartheta_e$ was defined by the computer simulation of the electron pass through the crystal with statistics $N = 5000$. The angular distribution of an electron radiation was calculated by formula (7). The sum was normalized to one incident electron.

For electron energy $\varepsilon = 1$ GeV the value of the Lindhard angle for a silicon crystal oriented by <111> axis to the electron beam direction is $\psi_L = 0.41$ mrad that is very close to the character value of the photon emission angle $\gamma^{-1} \approx 0.51$ mrad. Taking into account that the character value of electron scattering angle is close to $\psi_L$ (see Fig.2a, left) one can conclude that there is a dipole regime of radiation in this case. In fact, the angular distribution of radiation in Fig.2a (middle) shows the angular distribution of emitted photons, which is characteristic for dipole regime of radiation with one maximum near the crystal axis. The integral degree of linear polarization of this radiation is about 5% only, and the direction of polarization is vertical, i.e. along the y-axis (see Fig.2a, right).

The results of analogous calculations for 100 GeV electron beam in the silicon crystal of the same thickness and orientation are presented in Fig.2b. The initial beam divergence was taken as 1 μrad that is about five times less than $\gamma^{-1}$. One can see that the 10 μm crystal thickness is not enough to close the ring of coherent scattering for 100 GeV electrons (Fig.2b, left). There is also markedly weaker influence of incoherent scattering to the formation of angular distribution of scattered electrons than in the previous case ($\varepsilon = 1$ GeV). However, the most significant changes occur in angular distributions of radiation and polarization (Fig.2b, middle and right). In particular, for 100 GeV electrons case there is a depletion of the photon angular distribution at $\gamma\theta_x \approx 0$ and $\psi$ belonging to the interval $[\psi - \frac{1}{2}\gamma^{-1}, \psi + \frac{1}{2}\gamma^{-1}]$. The reason of these transformations is the essentially non-dipole regime of radiation in crystal for 100 GeV electrons: $\vartheta_e \approx \psi_L = 41$ μrad and $\gamma^{-1} \approx 5.1$ μrad, so, the non-dipole parameter is $\gamma\vartheta_e \approx 8$.

The integral degree of linear polarization is about 25 %, and it is horizontal (along x-axis) in this case. Using the slit-type horizontal photon collimator with the angular width $\Delta\theta = \gamma^{-1}$ and putting them as it sowed in Fig.2b (right) by dashed lines it is possible to obtain linearly polarized photon beam with polarization degree about 60 %.

Note, that just the special type of coherent scattering in a crystal and the non-dipole regime of radiation allow us to obtain a high degree linearly polarized photon beams on the basis the angular separation of emitted gamma-quanta.

The condition for the non-dipole regime of radiation, namely $\gamma\vartheta_e > 1$, can be fulfilled not only by increasing the electron energy but also using the crystals with higher atomic number.

The results of our calculations of scattering and radiation of 100 GeV electrons passed through the 10 μm tungsten crystal at the angle $\psi = \psi_L$ to the crystal axis <111> are presented in Fig. 3. The Lindhard angle in this case is $\psi_L = 125$ μrad that is equal to 24.4 in units $\gamma^{-1}$, so there is the condition of a deep non-dipole effect at radiation $\gamma\vartheta_e \gg 1$. The "doughnut scattering" effect takes place in this case also, and the whole angular distribution of scattered electron looks very similar to the electron distribution in Fig.2a (left) but in corresponding scale. In Fig. 3 (left) we can see only a small part of the whole angular distribution which is located around the initial direction of the incident electron beam. The angular distribution of emitted gamma-quanta presented in Fig. 3 (middle) demonstrates that despite a relatively wide angular distribution of scattered electron ($\Delta\vartheta_e \approx 2\psi_L = 250$ μrad) the radiation is concentrated mainly in a small cone ($\Delta\theta \approx 4\gamma^{-1} = 20$ μrad) along the incident electron beam direction with a "black hole" of $\gamma^{-1}$ in diameter in the middle of the cone.

The integral degree of linear polarization (horizontal) of radiation is about 5%. For the horizontal slit-type photon collimator with the angular width $\Delta\theta = \gamma^{-1}$ marked in Fig.3b (right) by dashed lines the degree of linear polarization (horizontal) of collimated photon beam comes up to 77 %. If we turn this collimator to the vertical position (in parallel to y-axis) we obtain the linearly (vertically) polarized photon beam with 70% of polarization degree. Note, that such a behavior of polarization is very similar to the amorphous target case but for the same efficiency of radiation it is necessary to use significantly thicker amorphous target.

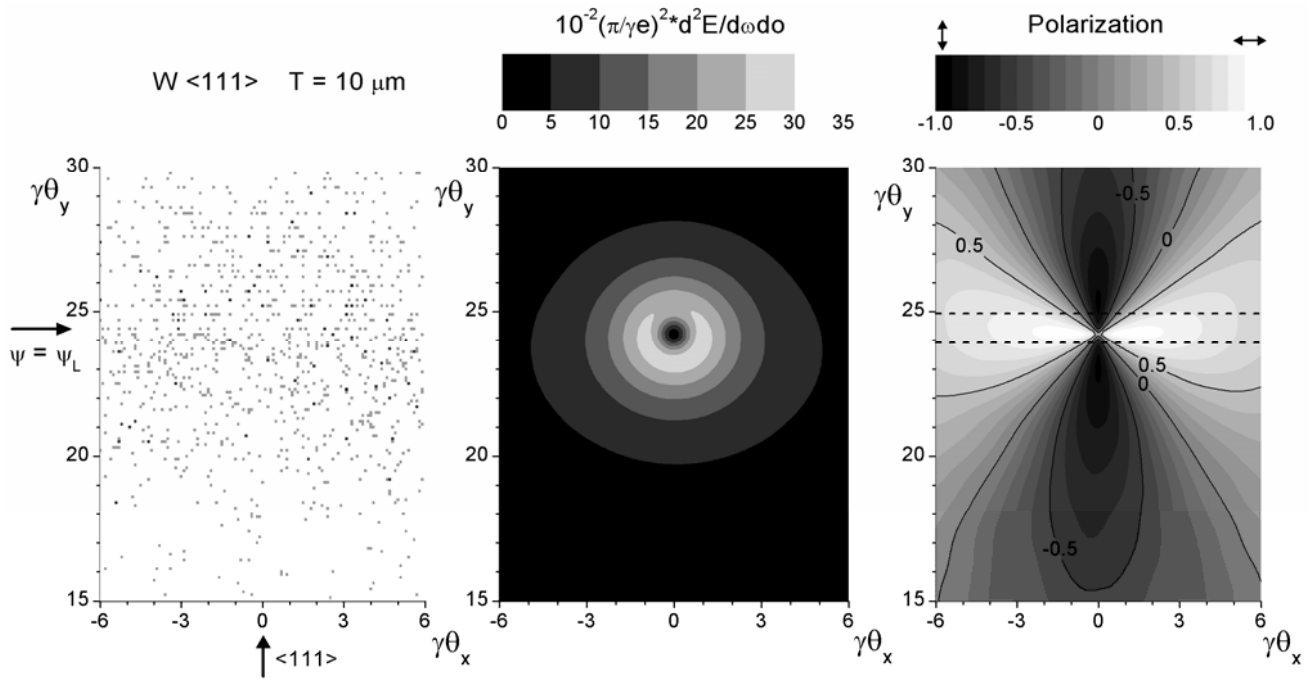

Fig. 3. The same as in Fig. 1 but for the tungsten crystal and 100 GeV of initial energy of the electron beam.

All these features of angular distributions of radiation and polarization are associated with the superposition of two different effects: coherent electron scattering by the crystal rows ("doughnut scattering" effect) and suppression of radiation due to this coherent multiple scattering (analogous to the LPM effect in an amorphous target).

## 4. CONCLUSION

- The present investigation shows a strong effect of the coherent multiple scattering of relativistic electrons in a thin crystal on the angular distributions of emitted gamma-quanta and their polarization.
- It is shown that circular polarization of each photon emitted in a thin crystal ($T \ll l_c$) is identically zero.
- The degree of linear polarization of collimated photon beam by the slit collimator with the angular width $\Delta\theta = \gamma^{-1}$ comes up to 77 % for 100 GeV electrons passed through the 10 μm tungsten crystal.
- For the experimental observation of the effects described above a high angular resolution (better than $\gamma^{-1}$) of the gamma-detector is needed, as well as a small (less than $\gamma^{-1}$) divergence of the electron beam.
- These effects must be taken into account when studying spectral-angular distributions of radiation and polarization by relativistic electrons in a crystal.